\documentclass[journal]{IEEEtran}

 \usepackage{xcolor}

\usepackage{ifpdf}

\usepackage{diagbox}

\usepackage{colortbl}
\definecolor{mygray}{gray}{.9}

\usepackage{cite}
\usepackage{multirow}

\usepackage[justification=centering]{caption}

\ifCLASSINFOpdf
  \usepackage[pdftex]{graphicx}
\else
\fi
\usepackage{subfigure}

\usepackage[ruled, linesnumbered]{algorithm2e}

\usepackage{algpseudocode}
\usepackage{amsmath}
\usepackage{enumitem}
\usepackage{amssymb}

\usepackage{array}
\begin{document}

\title{Dynamic Causal Attack Graph based Cyber-security Risk Assessment Framework for CTCS System}

\author{Zikai Zhang\\
Key Laboratory of Big Data \& Artificial Intelligence in Transportation (Beijing Jiaotong University), Ministry of Education, Beijing, China.\\
Email: zikaizhang@bjtu.edu.cn
}

\markboth{ IEEE XXX}%
{Zhang \MakeLowercase{\textit{et al.}}: }

\maketitle


\begin{abstract}
Protecting the security of the train control system is a critical issue to ensure the safe and reliable operation of high-speed trains. Scientific modeling and analysis for the security risk is a promising way to guarantee system security. However, the representation and assessment of the multi-staged, causally related, and temporal-dynamic changed attack dependencies are difficult in the train control system.
To solve the above challenges, a security assessment framework based on the Dynamical Causality Attack Graph (DCAG) model is introduced in this paper. Firstly, the DCAG model is generated based on the attack graph with consideration of temporal attack propagation and multi-stage attack event causality propagation. Then, the DCAG model is analyzed based on Bayesian inference and logic gateway-based inference. Through the case analysis of the CTCS-3 system, the security assessment framework is validated. With the DCAG-based security assessment framework, we can not only perform appropriate security risk quantification calculations, but also explore the importance of different attacks on system security risks, which is helpful in adjusting the cyber security defense policy.

\end{abstract}


\IEEEpeerreviewmaketitle

\section{Introduction}
\label{sec1}
Railways have been developing rapidly in the last few decades. Take China as an example, as of 2023,  the high-speed railway has a total length of 45 thousand kilometers, longer than a full circle around the equator of the earth.  As the vital equipment and technology of high-speed railway, the train control system is provided to control the spacing and speed of the train, to guarantee the safety, reliability, and efficiency of train operation \cite{10219050}.  The Chinese train control system (CTCS) is a large-scaled, multi-layered, and complicatedly structured train control system. To improve the automation and informatization level of CTCS systems, information technologies have been widely applied, which also bring great security risks that cause unnecessary braking, degraded running, emergency braking, and even collision accidents of trains \cite{9484083}.


Generally, a CTCS system can be viewed as a typical cyber-physical system, where the network and software work in the cyber domain and trains operate in the physical domain. So, the goal for our cyber security is to prevent the CTCS system from every malware attack and network attack\cite{wang2021survey}. To ensure cyber security, firewalls, anti-virus software, anomaly detection/intrusion detection technologies, and other cyber security protection methods are utilized on the CTCS system. However, due to the delay-sensitive character of the CTCS system, the security protection performances must be constrained by system functional requirements to ensure availability and efficiency\cite{8686209}. Moreover, with the improvement of attack technology and the discovery of new vulnerabilities, it is more difficult to protect the CTCS system.  So, it is important to assess cyber security risks and provide answers about "Which kind of attack is more dangerous?" or "What is the weak point of the current security protection configuration?".

Traditionally, attack graph models \cite{2019A2G2V} are widely used to describe attack scenarios with isolated vulnerabilities and attack paths. And the risk levels of the system are assessed utilizing Common Vulnerabilities and Exposures (CVE) \cite{Yosifova2021VulnerabilityTP} or Common Vulnerability Scoring System (CVSS) \cite{4042667}. However, directly assessing the security risk of CTCS system with the attack graph model is challenging for the following reasons. 
The first challenge is about the selection of the security risk metric to quantitatively measure the influence of a security-relevant attribute in the attack scenarios of the system. Just considering the vulnerability-related metric is not suitable, since some viable solutions may be deployed in the system to deal with the vulnerabilities. 
The second challenge is about the modeling of complex attack scenarios. Simply using an attack graph can not properly describe the multi-stage attacks (focus on risk spread probability), the causal-related attack propagations (focus on risk event trigger probability), and the temporal accumulation of security risks. 
The third challenge is about dealing with dynamic features in attack-defend scenarios. In the attack-defend process, vulnerabilities are discovered and patches are applied, novel attacks are designed, new protection techniques are deployed, and some attacks succeed and propagate in the system over time. 

To address the above challenges, a Dynamic Causal Attack Graph (DCAG) based cyber security risk assessment framework is introduced for CTCS systems.   Firstly, the DCAG model is generated based on the attack graph, with consideration of attack propagation and multi-stage attacks. And, during the generation process, the root node of the graph is converted from vulnerabilities to attack scenarios, the nodes from the previous time slice are added to deal with temporal security risk accumulation, and event causality relations are added on the edges to represent attack propagation dependencies. Then, combining Bayesian inference and logic gateway-based inference, a risk assessment method is proposed to deal with attack propagation and attack event causality propagation iteratively, which also provides the technical basis for the dynamic security risk assessment framework.

The contributions of this paper are as follows:
\begin{enumerate}
\item A DCAG modeling method is proposed to deal with complex attack scenarios considering temporal security risk accumulation, attack propagation, and multi-stage attack event causality propagation. 
\item Leveraging our DCAG modeling approach, a cyber security risk assessment process is provided to measure the influence of attack scenarios, combining Bayesian inference and logic gateway-based inference. 
\item A dynamic security risk assessment framework is proposed based on the DCAG model to evaluate the risk level of crucial components in the system. A case study based on CTCS-3 is illustrated to elucidate the validity of the proposed framework.
\end{enumerate}

The remainder of this paper is organized as follows:  
The preliminaries of this paper are introduced in Section II. The security risk assessment framework and the DCAG model are introduced in Section III. Section IV provides the case study, and Section V concludes this paper.

\section{Preliminary}
\subsection{Overview of CTCS-3 system }
\begin{figure}[htbp]
		\centering
		\setlength{\belowcaptionskip}{-1em}
		\includegraphics[scale=0.32]{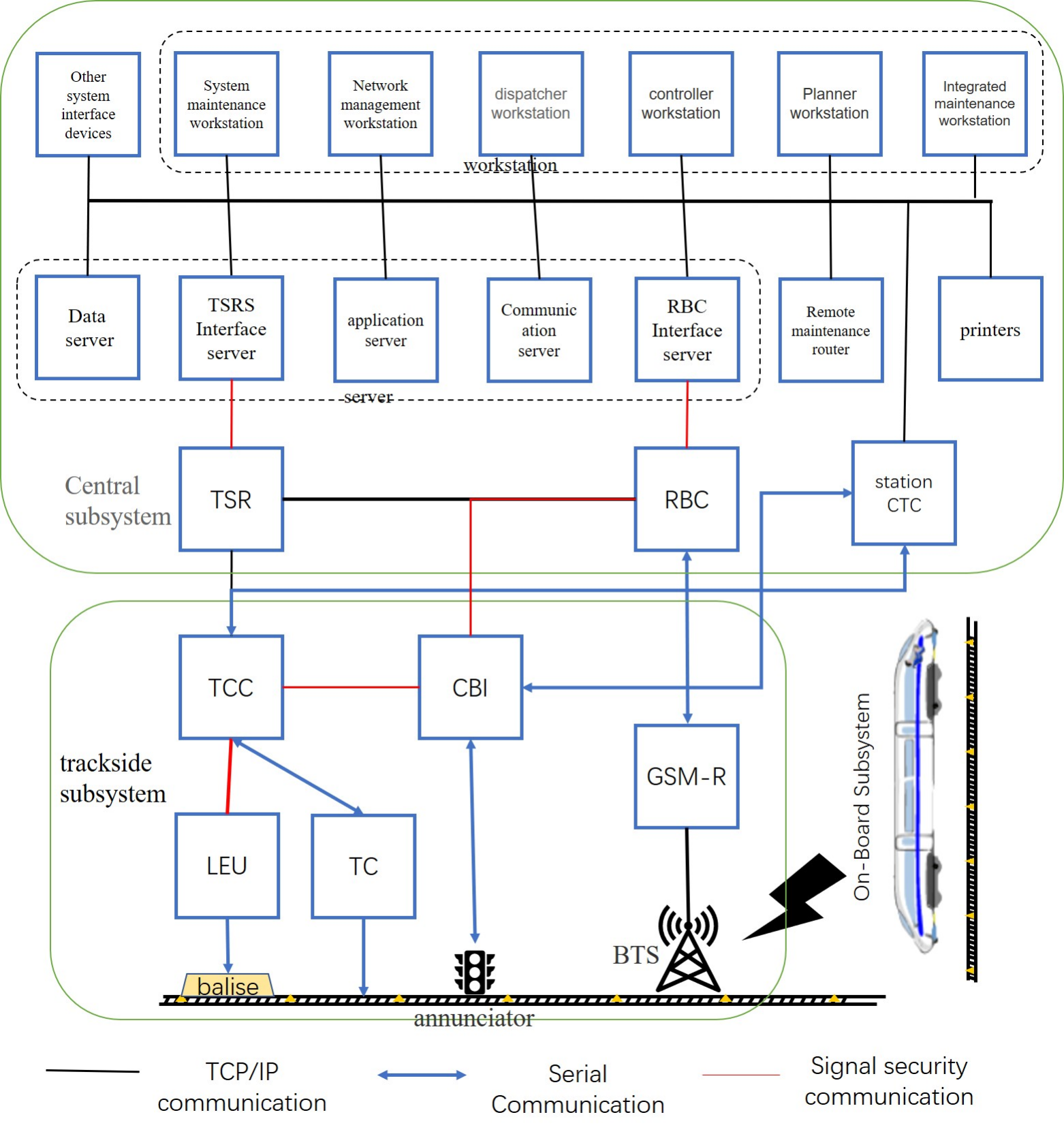}
		\caption{CTCS-3 train control system}
		\label{Fig.2}
	    \end{figure}

CTCS systems are safety-critical, and the fail-safe mechanisms are applied to achieve the demanded performance, including Reliability, Availability, Maintainability, and Safety (RAMS). As an example, the CTCS-3 train control system satisfies the requirements of a safety train operating with a speed range of 300-350 $km/h$ and minimal tracking intervals of 3min. The structure of CTCS-3 is shown in Fig. \ref{Fig.2}, including the on-board subsystem, the trackside subsystem, and the central subsystem. 

The on-board system is connected with the Base Transceiver Station (BTS) in the trackside subsystem based on Global System for Mobile Communication-Railway (GSM-R) wireless communication. 	 
On the trackside subsystem, the train control center (TCC) receives real-time train speed and position reports and track circuit (TC) occupancy information. The TC provides continuous train occupancy checking. Lineside Electronic Unit (LEU) and Balise systems are responsible for achieving the dynamic orientation of trains. The computer-based interlocking (CBI) subsystem is used as the basic signal equipment on the spot, which mainly controls the turnout and the annunciator in the station. 
On the central subsystem, the Centralized Traffic Control (CTC) system is the railway signaling technology equipment that conducts dispatchings and commands of the trains, and achieves centralized control by signal equipments. According to the occupation of the train within the  Radio Block Center(RBC)'s jurisdiction and the state of the incoming route, the Movement Authority (MA) is generated by the temporary speed limit information provided by CTC. Through GSM-R wireless communication, the MA is generated to ensure the safe operation of the train within the RBC's jurisdiction. Temporary Speed Restriction (TSR)  server is an important part of the third generation Chinese Train Contral System (CTCS-3) system. It provides the necessary speed limit management and control ability for the safe operation of trains.

While all subsystems of the CTCS-3 system (the on-board subsystem, the trackside subsystem, and the central subsystem) are susceptible to cyber-attacks in varying degrees, the central subsystem (IT) is particularly vulnerable due to its devices like switches, routers, and server interface that are connected and interaction with the Internet, leading to common malware and network attacks. And the devices in the trackside subsystem are connected with multiple communication protocols, as shown in Fig. \ref{Fig.2}, which leading to less network attacks. However, the devices in the trackside subsystem being physically exposed, lack of protection and potentially less secure than the central subsystem, can be targeted for exploitation with malware attack, wireless attack and risks propagated from the central subsystem, leading to disruptions in train operations, safety hazards, or even system failures. Additionally, the trackside subsystem's integration with the central subsystem and on-board subsystems, can create attack vectors that could compromise the entire CTCS-3 system.

The fail-safe mechanisms are applied in the CTCS system, where redundant and fault tolerance architectures are used to ensure its inherent safety, which at the same time bring some extra security protection capabilities. 
Examples are as follows. 
1) Railway system integrators and railway signal control enterprises are using SIL4 (Safety Integrity Level) safety computers. The software function independence principle of the safety computer makes the core software isolated from other software, which can effectively defeat malware attacks. 
2) The commonly used redundant voting mechanisms in CTCS system include the model of the 2-out-of-3 decision-making and the model of the double 2-vote-2 architecture. The redundant voting mechanism forces the attacker to solve the problem of attacking multiple types of targets under non-cooperative conditions coordinately. 
3) For different communication protocols, take the computer interlocking as an example, the master controller sends control instructions and receives status information through signal transmission and the conversion of TCP/IP Ethernet protocol into High-Level Data Link Control(HDLC) protocol. TCP/IP-based network attacks find it difficult to propagate to other communication protocols. Thus, the CTCS system is distinct from other systems in terms of assessing security risks.

CTCS systems usually consider functional safety with assessment methods based on IEC 62278 standard \cite{iec2002iec}. However, cyber security events are very different from safety events, which are often unexpected and sometimes involve incomplete information. Therefore, the security risk should be assessed under other guidelines for CTCS systems, such as ISO/IEC 27005 standard \cite{rajamaki2021resilience}, which defines guidelines related to information security. According to ISO/IEC 27005 standard, the security risk assessment of CTCS systems commences with identifying threats, followed by modeling and analyzing them.

\begin{table*}[]
\centering
\caption{The comparisons between related works and our method}
\begin{tabular}{lllll}
\hline
Challengings   & Methods & Features        & Weakpoints    & Our method        \\
\hline
\multirow{4}{*}{\begin{tabular}[c]{@{}l@{}}Dynamic\\ nature\\ of attacks\end{tabular}}        & \cite{8742636}      & \begin{tabular}[c]{@{}l@{}}Solving time-varying sequential \\characteristics with monte carlo\end{tabular}     & \begin{tabular}[c]{@{}l@{}}Focus on renewable variations stochastically \\ Other uncertainties are not captured\end{tabular} & \multirow{8}{*}{ \begin{tabular}[c]{@{}l@{}}A Dynamic\\ Causal Attack\\ Graph modeling\\ method considering \\temporal risk \\accumulation, attack \\propagation (risk \\spread probability),\\ and multi-stage \\attack event causality\\ propagation (risk \\event trigger \\probability)\end{tabular}} \\ 
 & \cite{Zhou2019PetrinetBA}      & Attack time is considered with Petri-net        & \begin{tabular}[c]{@{}l@{}}The evolution of attack states is ignored\end{tabular}       &                   \\
 & \cite{4591650}     & \begin{tabular}[c]{@{}l@{}}Using the dynamic Bayesian attack\\ graph for static and dynamic analysis\end{tabular}                               & \begin{tabular}[c]{@{}l@{}}The construction of attack graphs is NP-hard,\\the assumption of model is restrictive and static\end{tabular}   &                   \\
 & \cite{Sultan2017AMT}      & \begin{tabular}[c]{@{}l@{}}System behavior and attack actions are \\evaluated with model checking algorithms\end{tabular}             & \begin{tabular}[c]{@{}l@{}}The mutation operators of timed-automata are \\ not mathematically defined in the model\end{tabular}                            &                   \\
 \cline{1-4}
\multirow{4}{*}{\begin{tabular}[c]{@{}l@{}}Complexity\\ of attack\\ propagation\end{tabular}} & \cite{7446307}     & \begin{tabular}[c]{@{}l@{}}Multi-layered hierarchical Bayesian network to \\integrate static, dynamic, and behavior analysis\end{tabular}              & \begin{tabular}[c]{@{}l@{}}Only possible sequence of events without circles\\ can be inferred with backward reasoning.\end{tabular}                        &                   \\
 & \cite{gao2018exploring}    & \begin{tabular}[c]{@{}l@{}}Bayesian networks and probabilistic attack \\graphs for sophisticated multi-stage attacks\end{tabular} & \begin{tabular}[c]{@{}l@{}}The correlation between attacks is not considered \\and unable to model attack circles\end{tabular}                           &                   \\
 & \cite{Albanese2018AGM}     & \begin{tabular}[c]{@{}l@{}}Integrate attack and dependency graphs to \\assess the impact of multi-step attacks\end{tabular}                                              & \begin{tabular}[c]{@{}l@{}}Probability alone cannot solve how much risk\\ created and propagated by vulnerabilities together\end{tabular}   &                   \\
 & \cite{Frigault2017Measuring}     & \begin{tabular}[c]{@{}l@{}}Score-based probabilities are propagated with \\logic relations to cycles in the attack graph\end{tabular}          & \begin{tabular}[c]{@{}l@{}}The change of the security risk state over time is \\ not reflected in the model\end{tabular}                                    &     \\
 \hline
\end{tabular}
\label{tab:compare}	
\end{table*}

\subsection{Attack graph-based security risk assessment}
Security risk assessment is used to assess the vulnerability of threats and the resulting risk level of the system. A security risk assessment process is the starting point to protect the target system. It timely provides a view of the target system's cyber security state at a point, and aids to establish the cyber security defense policy.

As the most widely used technique to model attack scenarios, the attack graph model exhibits strong descriptive capabilities encompassing all potential attack pathways in the system \cite{Zeng2019Survey}. The basic attack graph model \cite{10.1007/978-3-319-39639-2_30,7437478,7446307} is a directed graph, which shows the possible attack sequence and attack effects. In the attack graph model,  edges are used to represent attack sequences, while vertices represent security-related elements such as hosts, services, vulnerabilities, permissions, or security states. 
Attack graphs are categorized into two types based on the roles or meanings of nodes). For the first type, such as the state enumeration attack graph \cite{Li2012NodeRankAA,10.1007/s10009-020-00594-9,9951035}, each node represents the system security state and edges represent state transitions caused by an attacker’s actions. For the second type, such as the dependency attack graph \cite{Fang2022BackPropagatingSD,9277665,akbarzadeh2023dependency}, each node represents the system security conditions in logic and edges indicate the causal relationships between the system conditions.

There are at least four types of attack graph analysis methods proposed to deal with system security risk assessment tasks, i.e., Bayesian network-based methods \cite{Joy2021PreciseEO,gao2018exploring,7446307}, Markov model-based methods \cite{Li2018DynamicSR,10207119,10284423}, graph-based methods \cite{MuozGonzlez2017EfficientAG,gao2018exploring,Albanese2018AGM}, and optimized methods \cite{Ghazo2019IdentificationOC,9763050,8742636,9619870}. These methods have specific characteristics \cite{Zeng2019Survey}. Graph-based methods perform better in terms of scalability. Bayesian
network-based methods offer significant advantages in solving ties and
correlation problems. Markov-based methods perform better in prediction.
Cost-optimized methods exhibit stronger portability.
However, these models just consider the system to be static and the dynamic nature of system properties is not taken into consideration. Vulnerabilities are regularly discovered and patched, while attacks can be detected by some detection methods or may result in the disconnection of some system components.
Incorporating these dynamic behaviors into the models is necessary to assess the system's cybersecurity risks. 
Some articles attempt to address this problem. In \cite{Zhou2019PetrinetBA}, the time required to execute an attack is considered in the attack graph, but the evolution of the system state is ignored. To track the ongoing attacks, a dynamic Bayesian attack graph model is proposed in \cite{4591650} for static and dynamic analysis. 
A Sequential Monte Carlo simulation method is proposed in \cite{8742636} to consider time-varying sequential characteristics. The evolution of the system security state can also be modeled with timed automata. In \cite{Sultan2017AMT}, the behavior of the system and the actions of the attacker are modeled and evaluated with model checking algorithms. 

However, the above solutions are not adapted to assess cyber security due to the complexity of attack propagation. Usually, an attack graph that is sufficiently large and detailed often contains cycles. To handle the complexity of the multi-stage attack with attack graph circles, some methods are proposed to cope with it. In \cite{Frigault2017Measuring}, CVSS base scores are converted into probabilities and then propagated along attack paths within an attack graph to obtain an overall metric, while giving special considerations and logic relations to cycles in the attack graph. In \cite{gao2018exploring}, the probabilistic attack graph is proposed according to the causal relationships among sophisticated multi-stage attacks by using Bayesian Networks. In \cite{7446307}, a
multi-layered hierarchical Bayesian network is proposed to integrate static analysis, dynamic analysis, and behavior
analysis for mobile systems. And, in \cite{Albanese2018AGM}, a mission-centric approach is proposed to integrate attack and dependency graphs to accurately assess the impact of multi-step attacks. This approach iteratively executes impact analysis and vulnerability remediation stages to perform a continuous system analysis.

In this paper, we propose the Dynamical Causality Attack Graph (DCAG) to model attack scenarios. Different from other methods, the DCAG-based method is particularly suitable for the CTCS system with two reasons. Firstly, as shown in Table I, the proposed DCAG-based method can deal with dynamic nature of attacks and complexity of attack propagation simultaneously. And, the proposed DCAG-based method solved the specific weak points of  literature reviews listed in Table I for both challenging problems. 
Then, the graph structures are naturally suited to complex CTCS systems for network risk analysis. Logic gate structure operation is also very conducive to the probability combination analysis of different scenarios. So, the proposed DCAG-based method is particularly suitable for CTCS by graph based attack and event causality propagation with logic gateway-based Bayesian inference.

\section{The Dynamical Causality Attack Graph-based security risk assessment method}

The working flow of our method is shown in Fig. \ref{Fig.process}. Our method is a circular process, and majorly consists of four steps:
\begin{itemize}
\item First, to identify the critical assets and security states (system conditions, network conditions, security detection conditions and other security knowledge). 
\item Second, to model attacks by attack graph.
\item Third, to convert the attack graph into a Dynamical Causality Attack Graph (DCAG).
\item Fourth, to assess security risk with Bayesian inference and logic gateway based inference.
\end{itemize}
More detailed introductions of our method are illustrated as follows.

\subsection{Identify the critical assets and security states}

We consider the CTCS-3 system as shown in Fig. \ref{Fig.2}. The asset in this system consists of five main parts, i.e., remote servers, local hosts, printers, routers, and trackside computer-embedded devices.
Trackside computer-embedded devices are hardly facing TCP/IP-based security attacks, since they are connected with serial communication. Then, we don't consider them in the security assessment process.

Vulnerabilities are weaknesses in a system that can be exploited by attackers to gain unauthorized access or cause harm. Identifying and addressing these vulnerabilities is a fundamental aspect of security assessment. By automated scanning with scanners \footnote{https://trickest.com/solutions/vulnerability-scanning}, the initial vulnerabilities per asset are listed in Table \ref{tab:0}. These vulnerabilities may lead to a range of hazards, such as data breaches, Denial of service (DoS) attacks, unauthorized access, and malware infection. And these hazards may cause disruptions in train operations or even system failures for the overall railway signalling system. Take printers as a case, many experts and literature agree that its vulnerabilities do not pose a serious threat.However, in new network attack methods such as ATP attacks, printers are often used as attack intermediaries, providing a larger and more convenient path for the spread of risks.

\begin{figure}[htbp]
		\centering
		\includegraphics[scale=0.42]{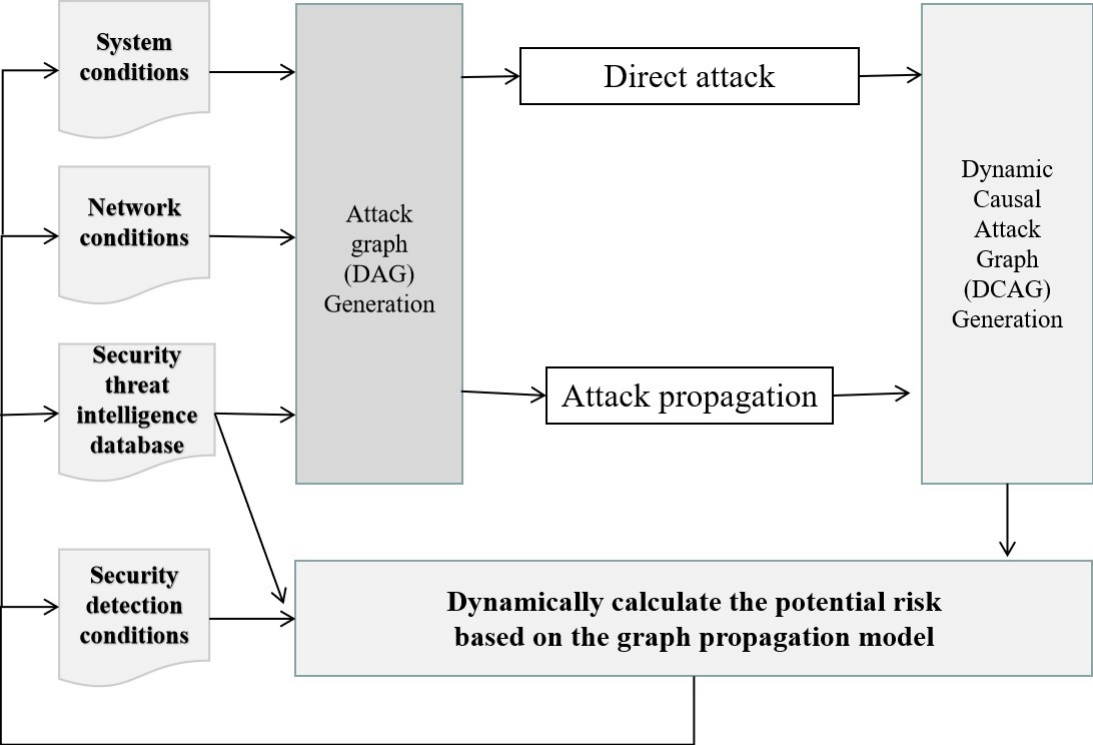}
		\caption{Architecture of our dynamic security risk assessment framework}
		\label{Fig.process}
\end{figure}

The security status of system assets used to represent by CVSS (https://www.first.org/cvss/) (Common Vulnerability Scoring System \cite{4042667}) scores. But, in this paper, the security states of the system are assessed based on the system condition (the system structure and the overall security states of system assets), the system network condition (the network traffic flow and the network protocol), the security detection condition (the recall rate and precision rate of detection) and the security threat intelligent database (the security threats, the threat severity, and the evolving attacks), as shown in Fig. \ref{Fig.process}. The higher the security risk of the whole system assessed, the higher the probability that system will be attacked.

\begin{table}
		\centering
		\caption{Assets and potential vulnerabilities}
		\begin{tabular}{p{0.2\columnwidth}| p{0.7\columnwidth}}
			\hline
			Asset &  Vulnerabilities   \\
			\hline
			Remote server & Remote procedure call vulnerability (CVE-2021-1664), Remote execution code vulnerability (CVE-2016-0118), Access control error vulnerability (CVE-2021-1674), Command injection Vulnerability (CVE-2020-0618), et al.\\
				\hline
			Local host  & Linux Denial of service vulnerability (CVE-2016-9919), Windows Denial of service vulnerability (CVE-2016-7237, CVE-2016-3369), Information leakage vulnerability of Linux (CVE-2013-3235), Information leakage vulnerability of Windows (CVE-2020-1033, CVE-2019-1202), Command injection Vulnerability (CVE-2020-0618), Cross-site scripting vulnerabilities (CVE-2021-23273, CVE-2021-27222), Office software Vulnerabilities (CVE-2017-11882, CVE-2018-0802), et al. \\	
				\hline
			Printer  & Printer Service Vulnerability (CVE-2021-1675, CVE-2020-1048), et al.  \\\hline
			Router  & Application hosting vulnerability (CVE-2019-1663), Privilege promotion vulnerability in authorization control (CVE-2020-3227), RCE remote execution code vulnerability (CVE-2020-3198), and Command injection Vulnerability (CVE-2020-3205), et al.
  \\	
			\hline
		\end{tabular}
		\label{tab:0}	
	\end{table}

\subsection{Attack graph generation}

 Attack graph generation focuses mostly on deriving dependencies between system vulnerabilities based on expert knowledge \cite{Salfer2018}. 
As depicted in Table \ref{tab:0}, there exist diversified possible attack scenarios with vulnerabilities in the target system environment.  And, the progressive attack behavior induces a transitive relationship between the vulnerabilities present in the system \cite{Dai2015ExploringRF}. 
Taking these facts into account, the attack graph emerges as a feasible approach to illustrate such cause-consequence relationships.

\begin{figure*}[!htbp]
		\centering
		\setlength{\belowcaptionskip}{-1em}
		\includegraphics[scale=0.24]{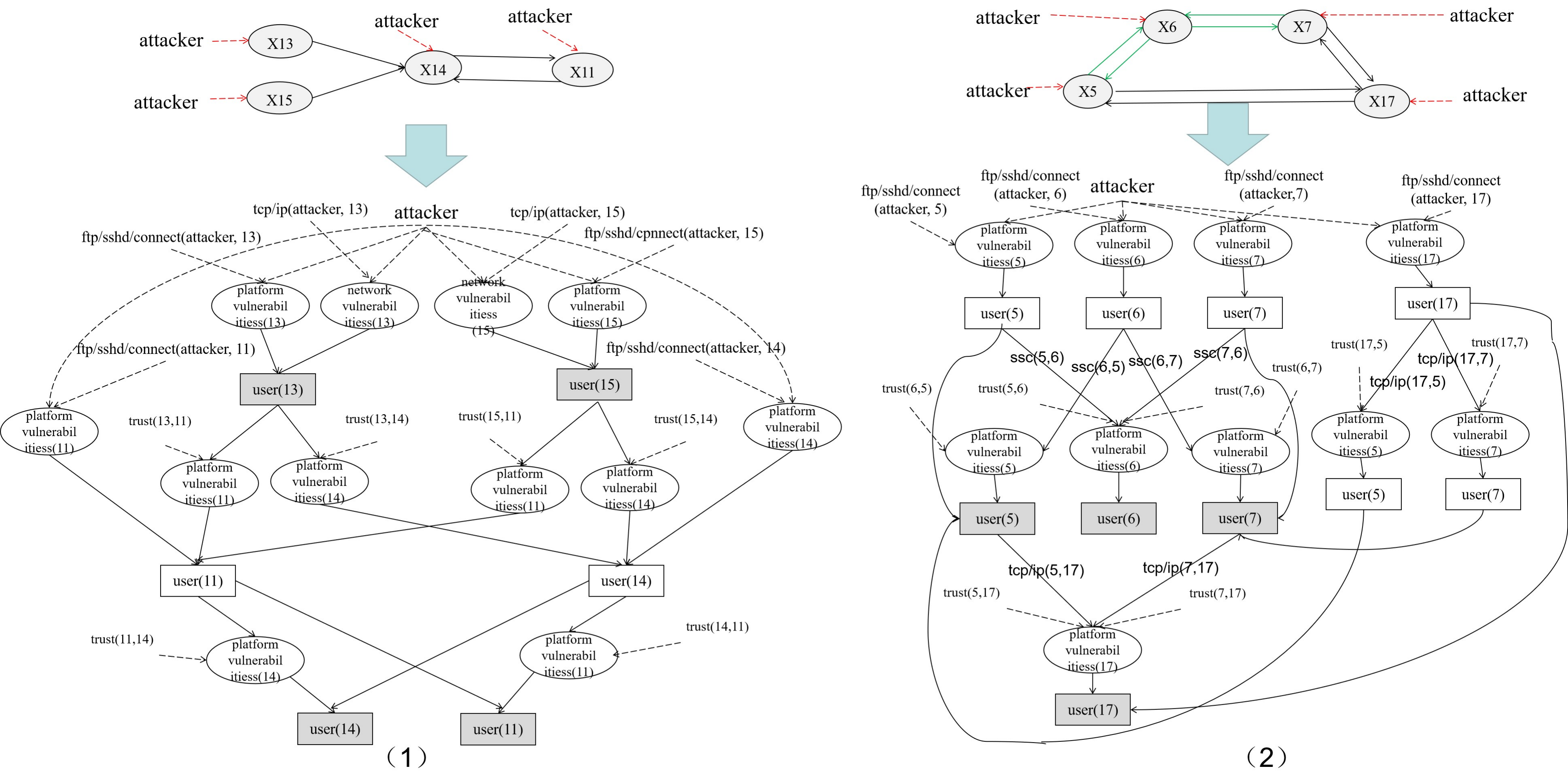}
		\caption{Samples of attack graph model}
		\label{Fig.example}
\end{figure*}

On the basis of the attack graph \cite{Zeng2019Survey}, two types of example scenarios can be modeled in Fig. \ref{Fig.example}. They depict a
clear picture of the possible attack scenarios for different system conditions. In Fig. \ref{Fig.example} (1), four nodes are subjected to malware attacks and network attacks due to their platform and network vulnerabilities. It can be seen that there exist inter-communications between node $X14$ and node $X11$. In Fig. \ref{Fig.example} (2), apart from the inter-communications, four nodes are connected with different protocols (presented by black and blue lines). Besides, these four nodes are facing malware attacks.

From Fig. \ref{Fig.example}, we can see that the attack graph can depict possible attack paths implicitly. However, there are some shortcomings. Firstly, it lacks the capability of characterizing the quantitative traits of threats. Second, it lacks the capability of measuring the propagation risk of the attack in the temporal domain. Thirdly, it also lacks the capability of dealing
with the causal relationships among different nodes. In consideration of these shortcomings, we turn to convert the attack graph to a dynamical causality attack graph to perform the security risk assessment.

\begin{figure*}[!htbp]
		\centering
		\setlength{\belowcaptionskip}{-1em}
		\includegraphics[scale=0.33]{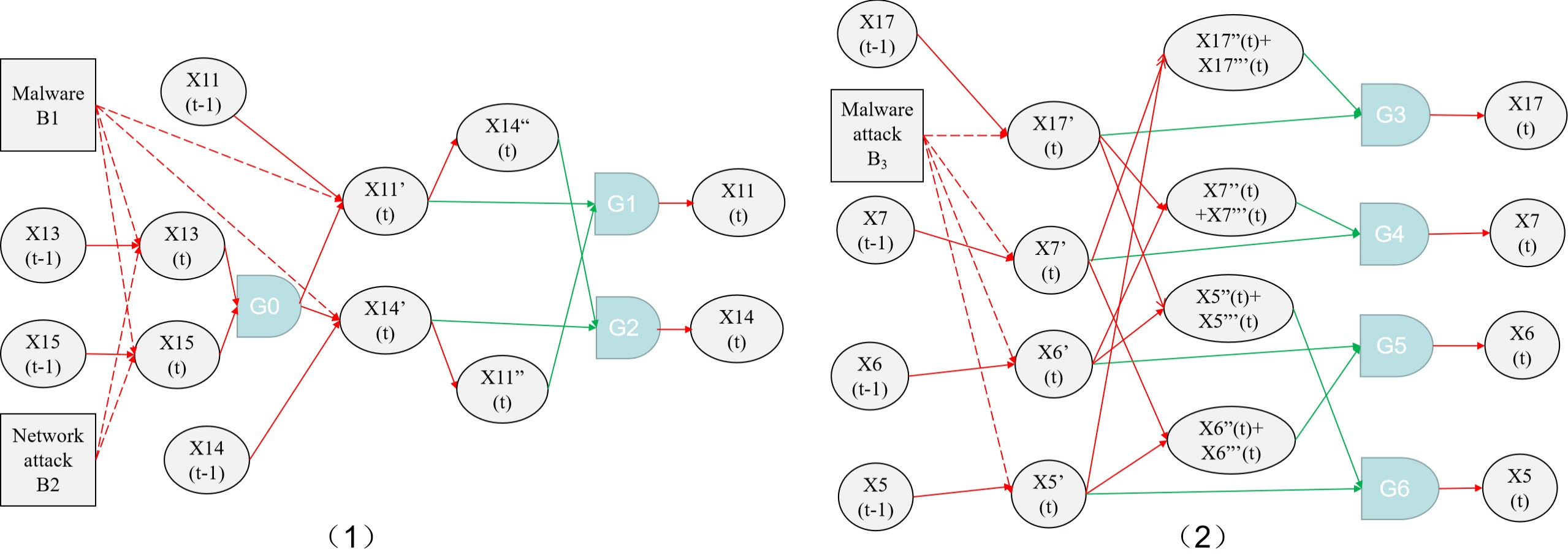}
		\caption{Samples of DCAG model generation based on attack graph}
		\label{Fig.convertexample}
\end{figure*}
\subsection{DCAG generation based on attack graph}

Dynamical causality attack graph (DCAG) is generated as a directed graph for causality representation. Nodes in DCAG model have three types, i.e., root nodes (Squares $B_i$),  exploit nodes and extended exploit nodes (Circles $X_i$),  and the logic gateway (traditional distinctive shape $G_i$). The root node represents the basic security risk. In practice, the security threats in a certain system usually require the combination of multiple attacks using different vulnerabilities. So, we just demonstrate the attack scenarios as root nodes, rather than the vulnerabilities. 
The exploit node represents the security risk of the corresponding asset or the security risk caused by other assets with attack propagation events. And, the logic gateway defines the combinational causality. For security risk propagation, the solid green arrow represents the linkage attack event matrix $P_{n;i}$ between $V_i$ (the parent variable of $X_n$) and $X_n$. The dotted red arrow represents the conditional attack event, and denoted as $A_{i,k;j,n}$, with the parameters $a_{n,k;i,j} \equiv Pr\{A_{n,k;i,j}\}$. The solid red arrow represents the weighted attack event matrix $F_{n,k;i,j}$ between $V_i$ and $X_n$. $F_{n,k;i,j} \equiv (r_{n,i}/r_n)A_{n,k;i,j}$, $r_{n,i}$ is the causal intensity between $X_n$ and $V_i$, $r_(n,i) \ge 0$ and $r_n=\sum_{i} r_{n,i}$.
For a child attack event $X_{n,k}$, the sum of causal influences from
all its parent attack events, and $X_{n,k} = \sum_{i} (r_{n,i}/r_n) \sum_{j}A_{n,k;i,j} V_{i,j}$. Then, the probabilistic reasoning of DCAG is calculated as,
\begin{align}
\operatorname{Pr}\left\{X_{n,k} \mid \bigcap_{i} V_{i,j_i}\right\}=\sum_{i}\left(r_{n ; i} / r_{n}\right) a_{n,k ; i,j} 
\end{align}

\begin{equation}
\begin{split}
x_{n,k} &\equiv \operatorname{Pr}\left\{X_{n,k}\right\} \\
&=\sum_{i}\left(r_{n ; i} / r_{n}\right) \sum_{j} a_{n,k ;i,{j}_{i}} \operatorname{Pr}\left\{ V_{i,j_i} \right\} \\
&=\sum_{i}\left(r_{n ; i} / r_{n}\right) \sum_{j_{i}} a_{n,k;i, j_{i}} v_{i,j_i .}
\end{split}
\end{equation}
Applying the equations above, it is easy to assess the security risk probability of every node under multi-stage attacks. 

To gain a more concrete understanding of the equations, we now consider a specific example. Supposing a network with two critical servers is vulnerable to a zero-day attack. The causal intensity (Risk Event Trigger Probability) represents the likelihood that a specific event triggers the attack. Let's assume there's a high chance (80\%) of this happening due to the growing black market for zero-day exploits. The linkage attack event probability (risk propagation rate) represents how likely the attack spread to other components in the network if it successfully infiltrates the server. We'll assume a 70\% chance of this example. The high trigger probability signifies a significant concern, and the spread probability indicates a potential domino effect. When assessing the overall risk, we need to consider them together. When set the risk level as 1, the risk score of the network is calculated as $2* risk\ propagation\ rate * causal\ intensity = 1.12$. By combining risk spread and trigger probability with risk level, a risk assessment becomes more accurate and helps prioritize security measures.

The dynamic inference of DCAG is performed in a manner of cumulative aggregation of static attack inferences on sequential time slices. Based on the Markov process, the DCAG attack events are modeled with two assessment time slots (i.e., time slot ‘$t$’ and time slot ‘$t-1$’). Then, this reasoning process could reflect the temporal correlations among attack events by representing the sequential causality interactions in risk propagating processes.

We can convert the attack graph to DCAG following five steps:
\begin{itemize}
\item[1.] Replace vulnerabilities to attack types as the attack root node.
\item[2.] Convert the attack graph to a Bayes-based attack graph.
\item[3.] Add the temporal correlation node to the graph.
\item[4.] Add the conditional function of each edge.
\item[5.] Add logic gateways with combinational causalities on the graph.
\end{itemize}

Based on the above conversion steps, Fig. \ref{Fig.convertexample} gives the DCAG of our example scenarios. As shown in Fig. \ref{Fig.convertexample}, we divide the attack scenarios into two types, i.e., network attacks and malware attacks. There are two types of gateways used in the example scenarios. Take Fig. \ref{Fig.convertexample} (1) as an example, gateway $G0$ represents the causality conditional sum-up correlation for nodes $X13$ and $X15$. Gateways $\{G1, ..., G6\}$ representation sum up correlations of the propagation risks. We set the security state $1$ represents the occurrence of the attack event and $2$ is the benign event. For logic gateway $G0$, when the security state is $1$, the state expression is $A_{0,1;13,1}X_{13,1}+A_{0,2;13,1}X_{13,2}A_{0,1;15,1}X_{15,1}$; when state is $2$, the state expression is  $A_{0,2;13,2}X_{13,2}A_{0,2;15,2}X_{15,2}$.
For logic gateway $G_{1-6}$, when the state is $1$, the state expression is $\sum_{i} X_{i,1}$; when the state is $2$, the state expression is  $\sum_{i} X_{i,2}$.
For extended exploit nodes, take $X11$ as an example, $X_{11",1}^{t}$ and $X_{11',1}^{t}$ are extended exploit nodes. The equation and inference procedure are as follows, 
\begin{equation}
\begin{split}
X_{11',1}^{t}=& \left(\mathrm{r}_{11 ; 14} /r_{11}\right) X_{14,1}^{t} A_{11,1 ; 14, 1}\\
&+\left(r_{11 ; 1} / r_{11}\right) X_{11,2}^{t-1} A_{11,1 ; 1,1} B_{1,1} \\
&+\left(r_{11 ; 2} / r_{11}\right) X_{11,2}^{t-1} A_{11,1 ; 2,1} G_{0,1}
\end{split}
\end{equation}
where the first line of the equation represents the risk propagated by node $X14$, the second line of the equation represents the risk propagated by a direct malware attack, and the last line of the equation represents the risk propagated by gateway $G0$ (i.e., node $X13$ and node $X15$).
\begin{equation}
X_{11",1}^{t}=\left(r_{11 ; 11_{t-1}} /r_{11}\right) X_{11,1}^{t-1} A_{11,1 ; 11_{t-1}, 1} 
\end{equation} 
which represents the risk of self-propagation over time.

When DCAG is generated, we can assess the security risk of the whole nodes by propagating the previous security risks and the current attack threats on the DCAG along with time.
From the example scenarios in Fig. \ref{Fig.example} and Fig. \ref{Fig.convertexample}, there are three differences compared with previous attack graphs. Firstly, DCAG uses nodes to represent dynamical temporal relations between time slots.  Second, DCAG uses extended nodes to represent the sub-attack events. Third, the causal relations between multi-stage attack events are established on edge.

\section{Case study: security risk assessment for CTCS-3 system}

\subsection{The modeling phase}
To assess the security risk of CTCS-3 system,  we first consider the inherent attributes of CTCS-3 system in the modeling process. 
Firstly, the balise, the annunciator, and the TC are ignored as security assets, since the cybersecurity risk in these components is relatively small due to the serial communication. And, when identifying the attack paths, the propagations of security risk through serial communication are ignored. Then, the attack paths between different communication protocols are ignored since it is not easy to occur. Finally, in order to simplify the modeling and calculation process, servers and workstations are merged into a large category for the overall evaluation.

\begin{figure*}[htbp]
		\centering
		\setlength{\belowcaptionskip}{-1em}
		\includegraphics[scale=0.32]{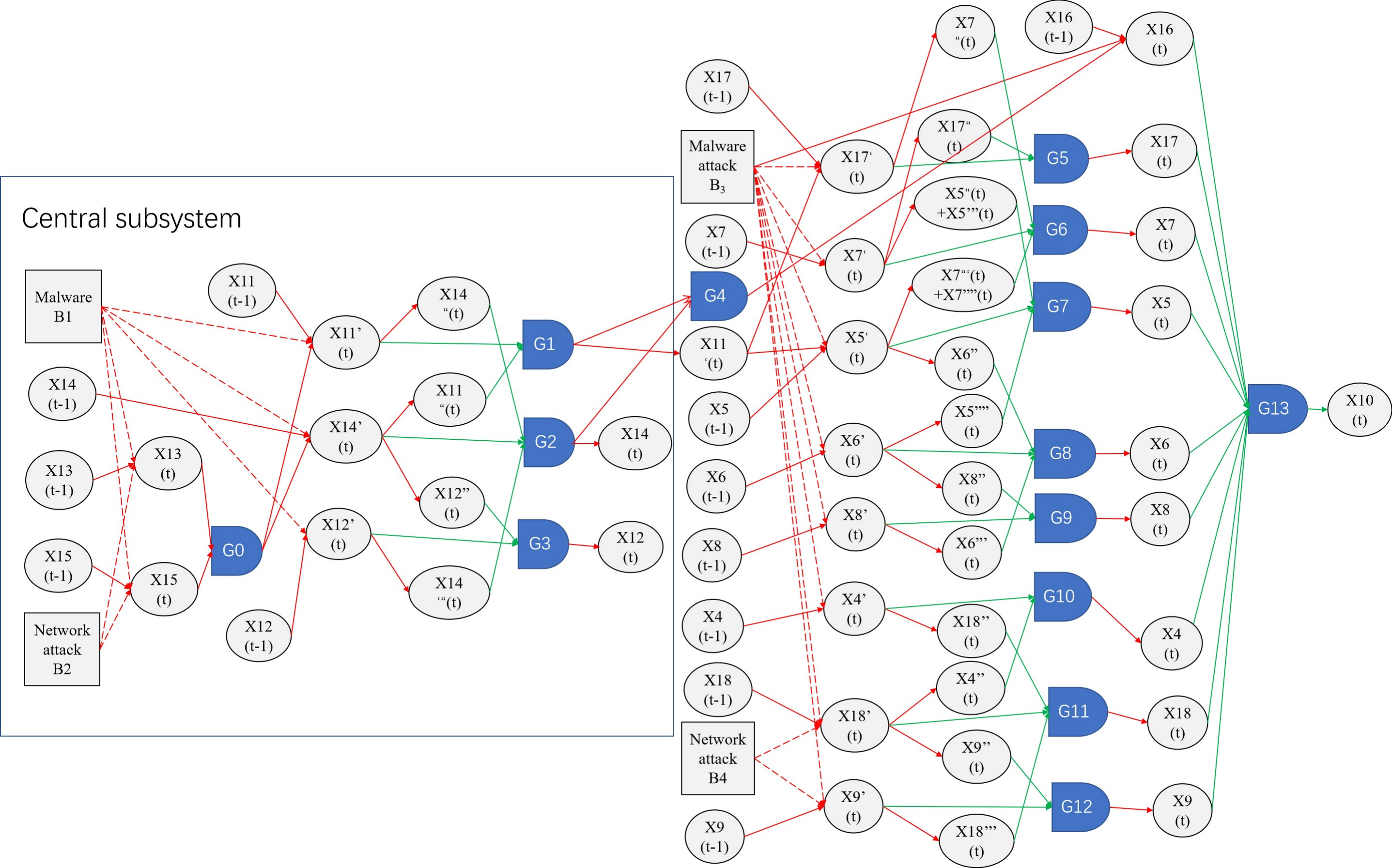}
		\caption{DCAG of CTCS-3 train control system}
		\label{Fig.DCAG}
	    \end{figure*}

The DCAG of the train control system is built as shown in Fig. \ref{Fig.DCAG}. The left side is the DCAG of the central subsystem with IT techniques, and the right side is the DCAG of the operational subsystem (the trackside subsystem) with industry control system (ICS) techniques. As shown in Fig. \ref{Fig.DCAG}, malware attacks and network attacks in different subsystems are distinguished individually, since the security risk level and risk types in different subsystems are different.
To describe the system more succinctly, all risk events are represented by symbols, as shown in Table I. Specially, for logic gateway $G0$, when the security state is $1$, the state expression is $A_{G0,1;13,1}X_{13,1}+A_{G0,2;13,1}X_{13,2}A_{G0,1;15,1}X_{15,1}$; when the security state is $2$, the state expression is  $A_{G0,2;13,2}X_{13,2}A_{G0,2;15,2}X_{15,2}$.  For logic gateway $G4$, when the security state is $1$, the state expression is $A_{G4,1;G1,1}X_{G1,1}+A_{G4,2;G1,1}X_{G1,2}A_{G4,1;G2,1}X_{G2,1}$; when the security state is $2$, the state expression is  $A_{G4,2;G1,2}X_{G1,2}A_{G4,2;G2,2}X_{G2,2}$.
For other logic gateways, when the state is $1$, the state expression is $\sum_{i} X_{i,1}$; when the state is $2$, the state expression is  $\sum_{i} X_{i,2}$.

\begin{table}[tb]
    \centering
    \caption{Descriptions of the DCAG symbols}
    \begin{tabular}{ll} 
    \hline
			Node Id & Node Definition   \\
			\hline
			$G(0-13)$ & logic gateways       \\	
			 $B_1$  & The malware attack in IT environment   \\		
			 $B_2$  & The network attack in IT environment \\	
			$B_3$  & The malware attack in OT environment \\
			 $B_4$  & The network attack in OT environment\\
			 $X_4$  & BTS is at security risk or not  \\	
			  $X_5$  & RBC is at security risk or not  \\	
			   $X_6$  & CBI is at security risk or not  \\	
			    $X_7$  & TCC is at security risk or not  \\	
			     $X_8$  & LEU is at security risk or not  \\	
			      $X_9$  & On-board system is at security risk or not  \\
			       $X_{10}$  & CTCS-3 system is at security risk or not  \\
			        $X_{11}$  & Server in CTC system is at security risk \\ & or not  \\	
			         $X_{12}$  & Printer in CTC system is at security risk \\ & or not  \\
			          $X_{13}$  & Remote maintenance router in CTC system \\ & is at security risk or not  \\
			           $X_{14}$  & Workstation in CTC system is at security \\ &risk or not  \\
			            $X_{15}$  & Other system interface device in CTC \\ & system is at security risk or not  \\
			             $X_{16}$  & Station CTC is at security risk or not \\
			              $X_{17}$  & TSR is at security risk or not  \\
			              $X_{18}$  & GSM-R is at security risk or not  \\
			\hline
		\end{tabular}
		\label{tab:1}	
	\end{table}	    

\subsection{The propagation phase}
According to the algorithm of DCAGs, the formulas to calculate every node’s occurrence probability are shown below, where 1 represents the occurrence of the event and 2 is the opposite:
\begin{equation}
\begin{split}
X_{13,1}^{t}=&\left(r_{13 ; B1} / r_{13}\right) X_{13,2}^{t-1} A_{13,1 ; B1,1} B_{1,1} \\
&+\left(r_{13 ; B2} / r_{13}\right) X_{13,2}^{t-1} A_{13,1 ; B2,1} B_{2,1}\\
&+\left(r_{13 ; 13_{t-1}} /r_{13}\right) X_{13,1}^{t-1} A_{13,1 ; 13_{t-1}, 1} 
\end{split}
\end{equation}

\begin{equation}
\begin{split}
X_{15,1}^{t}=&\left(r_{15 ; B1} / r_{15}\right) X_{15,2}^{t-1} A_{15,1 ; B1,1} B_{1,1}\\
&+\left(r_{15 ; B2} / r_{15}\right) X_{15,2}^{t-1} A_{15,1 ; B2,1} B_{2,1}\\
&+\left(\mathrm{r}_{15 ; 15_{t-1}} /r_{15}\right) X_{15,1}^{t-1} A_{15,1 ; 15_{t-1}, 1}
\end{split}
\end{equation}

\begin{equation}
\begin{split}
X_{11,1}^{t}=& X_{11",1}^{t} + X_{11',1}^{t} \\
=&\left(\mathrm{r}_{11 ; 14} /r_{11}\right) X_{14,1}^{t} A_{11,1 ; 14, 1}\\
&+\left(r_{11 ; B1} / r_{11}\right) X_{11,2}^{t-1} A_{11,1 ; B1,1} B_{1,1} \\
&+\left(r_{11 ; G0} / r_{11}\right) X_{11,2}^{t-1} A_{11,1 ; G0,1} G_{0,1}\\
&+\left(r_{11 ; 11_{t-1}} /r_{11}\right) X_{11,1}^{t-1} A_{11,1 ; 11_{t-1}, 1} 
\end{split}
\end{equation}

\begin{equation}
\begin{split}
X_{14,1}^{t}=& X_{14"',1}^{t}+X_{14",1}^{t} + X_{14',1}^{t}\\
=& \left(\mathrm{r}_{14 ; 12} /r_{14}\right) X_{12,1}^{t} A_{14,1 ; 12, 1} \\
&+ \left(\mathrm{r}_{14 ; 11} /r_{14}\right) X_{11,1}^{t} A_{14,1 ; 11, 1} \\
&+ \left(r_{14 ; B1} / r_{14}\right) X_{14,2}^{t-1} A_{14,1 ; B1,1} B_{1,1}\\
&+\left(r_{14 ; G0} / r_{14}\right) X_{14,2}^{t-1} A_{14,1 ; G0,1} G_{0,1}\\
&+\left(\mathrm{r}_{14 ; 14_{t-1}} /r_{14}\right) X_{14,1}^{t-1} A_{14,1 ; 14_{t-1}, 1}
\end{split}
\end{equation}

\begin{equation}
\begin{split}
X_{12,1}^{t}=& X_{12",1}^{t} + X_{12',1}^{t}\\
=& \left(\mathrm{r}_{12 ; 14} /r_{12}\right) X_{14,1}^{t} A_{12,1 ; 14, 1} \\
&+ \left(r_{12 ; B1} / r_{12}\right) X_{12,2}^{t-1} A_{12,1 ; B1,1} B_{1,1}\\
&+\left(\mathrm{r}_{12 ; 12_{t-1}} /r_{12}\right) X_{12,1}^{t-1} A_{12,1 ; 12_{t-1}, 1}
\end{split}
\end{equation}

\begin{equation}
\begin{split}
X_{7,1}^{t}=&X_{7"",1}^{t}+ X_{7"',1}^{t}+X_{7",1}^{t} + X_{7',1}^{t}\\
=& \left(\mathrm{r}_{7 ; \hat{5}} /r_{7}\right) X_{7,1}^{t} A_{7,1 ; \hat{5}, 1}\\
&+ \left(\mathrm{r}_{7 ; 5} /r_{7}\right) X_{7,1}^{t} A_{7,1 ; 5, 1}\\
&+ \left(\mathrm{r}_{7 ; 17} /r_{7}\right) X_{17,1}^{t} A_{7,1 ; 17, 1}\\
&+ \left(r_{7 ; B3} / r_{7}\right) X_{7,2}^{t-1} A_{7,1 ; B3,1} B_{3,1} \\
&+\left(r_{7 ; 7_{t-1}} / r_{7}\right) X_{7,1}^{t-1} A_{7,1 ; 7_{t-1}, 1}
\end{split}
\end{equation}

where ${r}_{7 ; \hat{5}}$ and $A_{7,1 ; \hat{5}, 1}$ represent the causal relation and the risk propagation probability between node $X5$ and node $X7$ connected with another communication protocol (i.e., signal security communication).
\begin{equation}
\begin{split}
X_{5,1}^{t}=&X_{5"",1}^{t}+ X_{5"',1}^{t}+X_{5",1}^{t} + X_{5',1}^{t}\\
=& \left(\mathrm{r}_{5 ; 6} /r_{5}\right) X_{6,1}^{t} A_{5,1 ; 6, 1}\\
&+ \left(\mathrm{r}_{5 ; \hat{7}} /r_{5}\right) X_{7,1}^{t} A_{5,1 ; \hat{7}, 1}\\
&+ \left(\mathrm{r}_{5 ; 7} /r_{5}\right) X_{7,1}^{t} A_{5,1 ; 7, 1}\\
&+ \left(r_{5 ; B3} / r_{5}\right) X_{5,2}^{t-1} A_{5,1 ; B3,1} B_{3,1} \\
&+ \left(r_{5 ; 11} / r_{5}\right) X_{5,2}^{t-1} A_{5,1 ; 11,1} X_{11,1}^{t} \\
&+\left(r_{5 ; 5_{t-1}} / r_{5}\right) X_{5,1}^{t-1} A_{5,1 ; 5_{t-1}, 1}
\end{split}
\end{equation}

where ${r}_{5 ; \hat{7}}$ and $A_{5,1 ; \hat{7}, 1}$ represent the causal relation and the risk propagation probability between node $X5$ and node $X7$ connected with another communication protocol (i.e., signal security communication).
\begin{equation}
\begin{split}
X_{16,1}^{t}=&\left(r_{16 ; B3} / r_{16}\right) X_{16,2}^{t-1} A_{16,1 ; B3,1} B_{3,1}
\\&+\left(r_{16 ; G4} / r_{16}\right) X_{16,2}^{t-1} A_{16,1 ; G4,1} G_{4,1}
\\&+\left(r_{16 ; 16_{t-1}} /r_{16}\right) X_{16,1}^{t-1} A_{16,1 ; 16_{t-1}, 1}
\end{split}
\end{equation}

\begin{equation}
\begin{split}
X_{6,1}^{t}=&X_{6"',1}^{t}+X_{6",1}^{t} + X_{6',1}^{t}\nonumber\\
=& \left(\mathrm{r}_{6 ; 8} /r_{6}\right) X_{8,1}^{t} A_{6,1 ; 8, 1}\\
&+ \left(\mathrm{r}_{6 ; 5} /r_{6}\right) X_{5,1}^{t} A_{6,1 ; 5, 1}\\
&+ \left(\mathrm{r}_{6 ; 3} /r_{6}\right) X_{6,2}^{t} A_{6,1 ; 3, 1}\\
&+\left(r_{6 ; 6_{t-1}} / r_{6}\right) X_{6,1}^{t-1} A_{6,1 ; 6_{t-1}, 1}
\end{split}
\end{equation}

\begin{equation}
\begin{split}
X_{18,1}^{t}=&X_{18"',1}^{t}+X_{18",1}^{t} + X_{18',1}^{t}\\
=& \left(\mathrm{r}_{18 ; 9} /r_{18}\right) X_{9,1}^{t} A_{18,1 ; 9, 1}\\
&+ \left(\mathrm{r}_{18 ; 4} /r_{18}\right) X_{4,1}^{t} A_{18,1 ; 4, 1}\\
&+ \left(\mathrm{r}_{18 ; B3} /r_{18}\right) X_{18,2}^{t} A_{18,1 ; B3, 1}B_{3,1}\\
&+ \left(\mathrm{r}_{18 ; B4} /r_{18}\right) X_{18,2}^{t} A_{18,1 ; B4, 1}B_{4,1}\\
&+\left(r_{18 ; 18_{t-1}} / r_{18}\right) X_{18,1}^{t-1} A_{18,1 ; 18_{t-1}, 1}
\end{split}
\end{equation}

\begin{equation}
\begin{split}
X_{17,1}^{t}=&X_{17",1}^{t} + X_{17',1}^{t}\\
=& \left(\mathrm{r}_{17 ; 7} /r_{17}\right) X_{7,1}^{t} A_{17,1 ; 7, 1} \\
&+ \left(\mathrm{r}_{17 ; B3} /r_{17}\right) X_{17,2}^{t} A_{17,1 ; B3, 1}B_{3,1}\\
&+ \left(\mathrm{r}_{17 ; 11} /r_{17}\right) X_{17,2}^{t} A_{17,1 ; 11, 1}X_{11,1}^{t}\\
&+\left(r_{17 ; 17_{t-1}} / r_{17}\right) X_{17,1}^{t-1} A_{17,1 ; 17_{t-1}, 1}
\end{split}
\end{equation}

\begin{equation}
\begin{split}
X_{4,1}^{t}=&X_{4",1}^{t} + X_{4',1}^{t}\\
=& \left(\mathrm{r}_{4 ; 18} /r_{4}\right) X_{18,1}^{t} A_{4,1 ; 18, 1} \\
&+ \left(\mathrm{r}_{4 ; B3} /r_{4}\right) X_{4,2}^{t} A_{4,1 ; B3, 1}B_{3,1}\\
&+\left(r_{4 ; 4_{t-1}} / r_{4}\right) X_{4,1}^{t-1} A_{4,1 ; 4_{t-1}, 1}
\end{split}
\end{equation}

\begin{equation}
\begin{split}
X_{8,1}^{t}=&X_{8",1}^{t} + X_{8',1}^{t}\\
=& \left(\mathrm{r}_{8 ; 6} /r_{8}\right) X_{6,1}^{t} A_{8,1 ; 6, 1} \\
&+ \left(\mathrm{r}_{8 ; B3} /r_{8}\right) X_{8,2}^{t} A_{8,1 ; B3, 1}B_{3,1}\\
&+\left(r_{8 ; 8_{t-1}} / r_{8}\right) X_{8,1}^{t-1} A_{8,1 ; 8_{t-1}, 1}
\end{split}
\end{equation}

\begin{equation}
\begin{split}
X_{9,1}^{t}=&X_{9",1}^{t} + X_{9',1}^{t}\\
=& \left(\mathrm{r}_{9 ; 18} /r_{9}\right) X_{18,1}^{t} A_{9,1 ; 18, 1} \\
&+ \left(\mathrm{r}_{9 ; B3} /r_{9}\right) X_{9,2}^{t} A_{9,1 ; B3, 1}B_{3,1}\\
&+ \left(\mathrm{r}_{9 ; B4} /r_{9}\right) X_{9,2}^{t} A_{9,1 ; B4, 1}B_{4,1}\\
&+\left(r_{9 ; 9_{t-1}} / r_{9}\right) X_{9,1}^{t-1} A_{9,1 ; 9_{t-1}, 1}
\end{split}
\end{equation}

\begin{equation}
\begin{split}
X_{10,1}^{t}=&1/9 (X_{9,1}^{t} +X_{18,1}^{t} +X_{4,1}^{t} +X_{8,1}^{t} + X_{6,1}^{t}\\
&+ X_{5,1}^{t} +X_{17,1}^{t} +X_{7,1}^{t} +X_{16,1}^{t})
\end{split}
\end{equation}

\subsection{The analysis phase}
In this case study, the security risk level of malware on central subsystem $P(B_{1,1})$  and trackside subsystem $P(B_{3,1})$ are set as 2. The security risk level of a networked attack on central subsystem $P(B_{2,1})$ is set as 1. And, the security risk level of wireless attack on trackside subsystem $P(B_{4,1})$ is set as 1.       

\begin{figure}[htbp]
		\centering
		\setlength{\belowcaptionskip}{-1em}
		\includegraphics[scale=0.30]{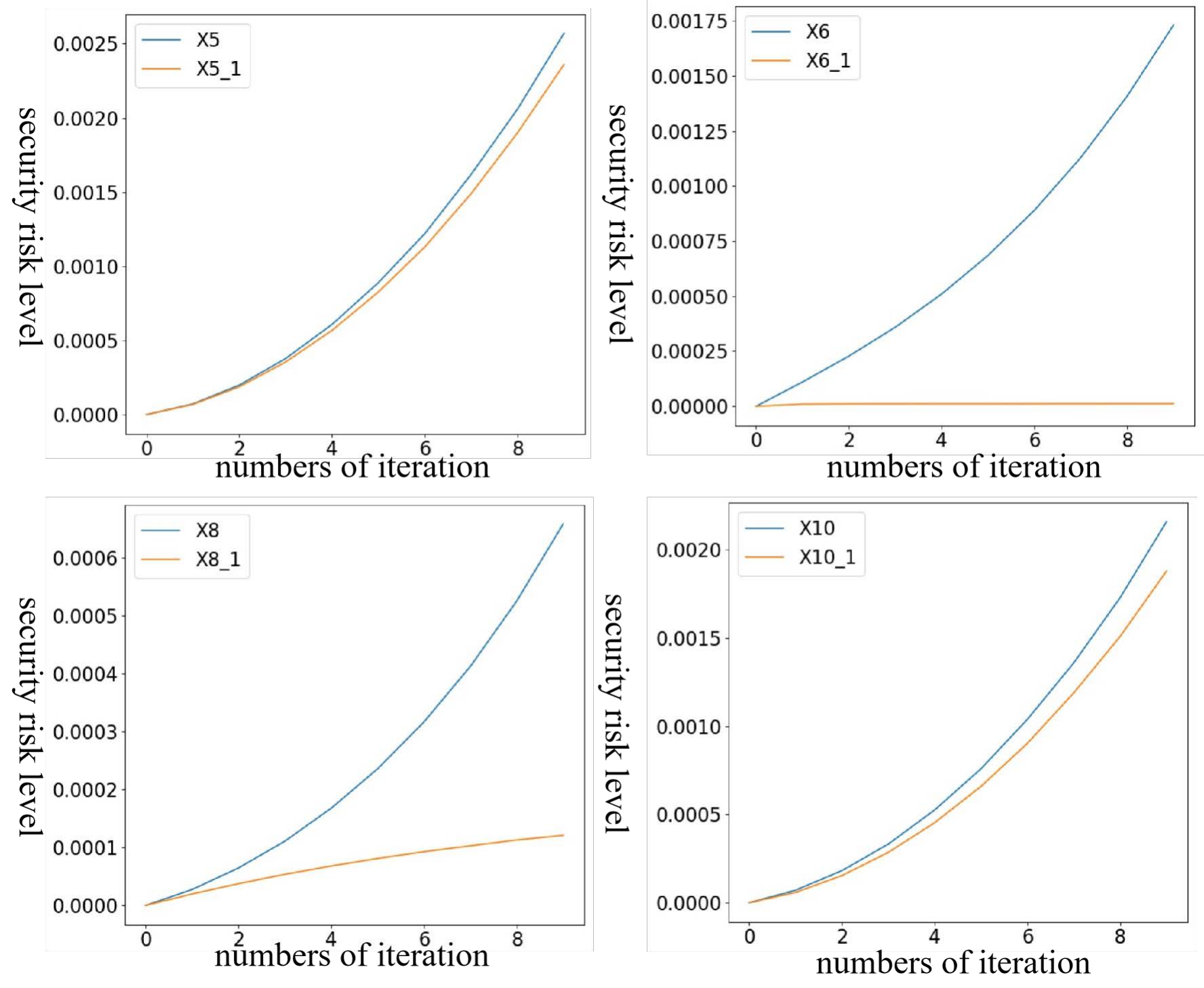}
		\caption{Risk propagation comparisons considering functional safety of CBI}
		\label{Fig.6}
	    \end{figure}
Usually, the attack initiation probability is determined by vulnerabilities  \cite{cai2020tiedao}. However, the above method does not consider the security risk self-propagation along with time. Moreover, only considering the vulnerabilities is not enough, the security risk propagation rates are also related to the anti-attack techniques and their own security attributes.
In this case study, the security risk self-propagation rate $P( A_{i,1 ; i_{t-1}, 1})$ is set as 0.9. 
For the gateway $G_0$, we assign each risk propagation probability as $0.01$. For the risk propagation probability in the central subsystem, we assign the risk propagation probability of malware risk and network attack risk as $0.001$ and $0.0001$, individually. For the risk propagation probability in the trackside subsystem, we assign the risk propagation probability of malware risk and network attack risk as $0.00001$ and $0.0001$, individually. For the inner risk propagation probability in the central subsystem, we assign the risk propagation probability as $0.5$. 
For the risk propagation probability in signal security communication, we assign the risk propagation probability as $0.001$.

We set the overall casual relations as 1 for each node in the graph. And, the traffic between nodes in the network can usually be related to risk causality in practice. In this paper, we assume that all communications are equally important. For nodes in $\{4,8,12,13,15,16\}$, we set the casual intensity between malware to them as 0.75. For nodes in $\{6,9,11,17,18\}$, we set the casual intensity between malware to them as 0.5. For nodes in $\{7,14\}$, we set the casual intensity between malware to them as 0.4. And, other casual relation probabilities are equally divided.

\begin{figure}[htbp]
		\centering
		\setlength{\belowcaptionskip}{-1em}
		\includegraphics[scale=0.28]{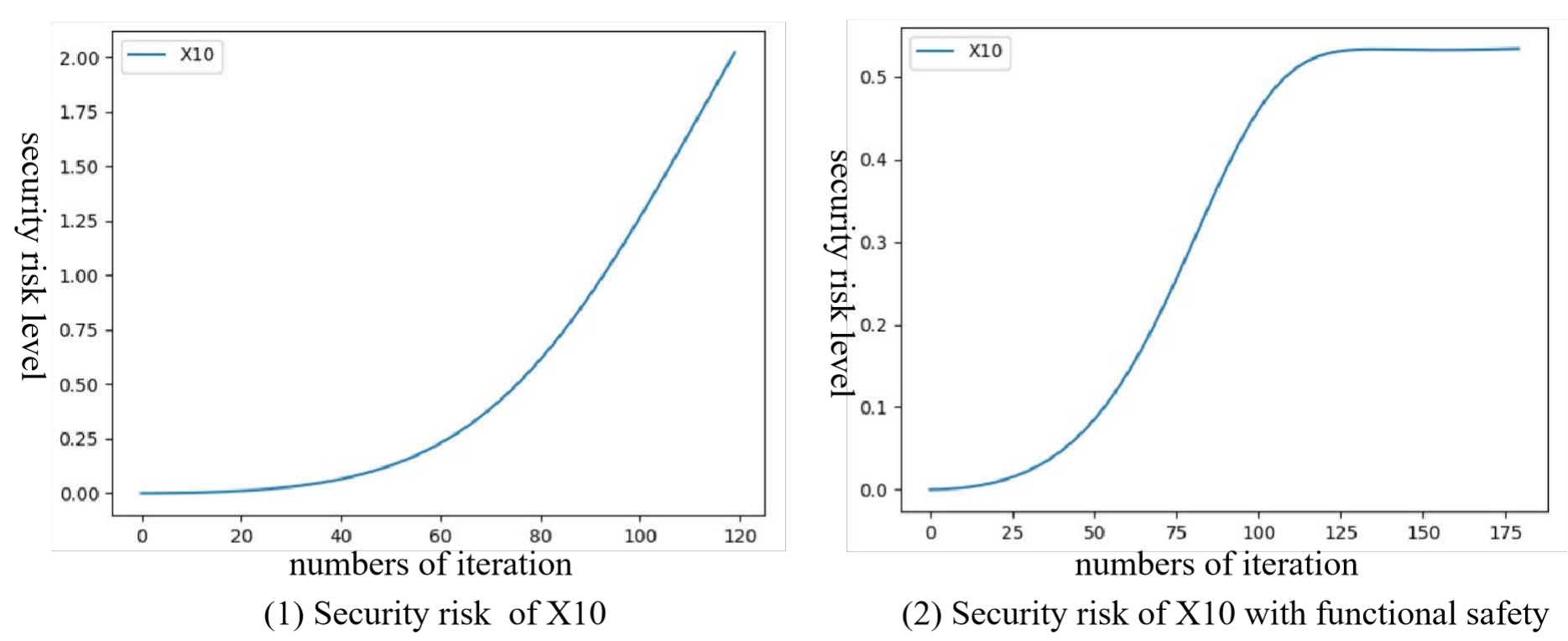}
		\caption{Long-term influence of risk propagation comparisons considering functional safety of CBI}
		\label{Fig.7}
	    \end{figure}
Firstly, considering the functional safety of CBI, we suppose it will be helpful to prevent security risk propagation, and assign the relevant risk propagation probability as $0.001$. Then, we compare the security risk propagation results of two scenarios based on considering the functional safety of CBI or not. The results are shown in Fig. \ref{Fig.6}, where $Xi$ denotes the scenario not considering the functional safety of CBI, and $Xi\_1$ denotes the scenario considering the functional safety of CBI. From Fig. \ref{Fig.6}, We can see that the security propagation is blocked slower with the functional safety of CBI. This assessment result reflects the real CTCS system characteristic of CBI to show the correctness of the proposed DCAG-based method.

We further compare these two scenarios with more risk propagation iterations. When the functional safety of CBI is not considered in the analysis, the propagation ability to security risks is improved, and the system is totally at security risk after 120 iterations, as shown in Fig. \ref{Fig.7}. However, when considering the functional safety of CBI, the system security risk will be no more than 0.6. From Fig. \ref{Fig.7}, We can see that the functional safety of CBI has a long-term influence on security risk propagation. Based on the above simulations and observations, the proposed cyber security risk assessment method works well with CTCS-3 system features, and can effectively assess risk with sufficient resolution.
More specifically, by measuring the risk level of the system, the proposed cyber security risk assessment method could give a metric to measure the benefits of applying or changing security technologies, security devices, and security strategies. And it will result in a increased secure system environment.

\begin{figure}[htbp]
		\centering
		\setlength{\belowcaptionskip}{-1em}
		\includegraphics[scale=0.32]{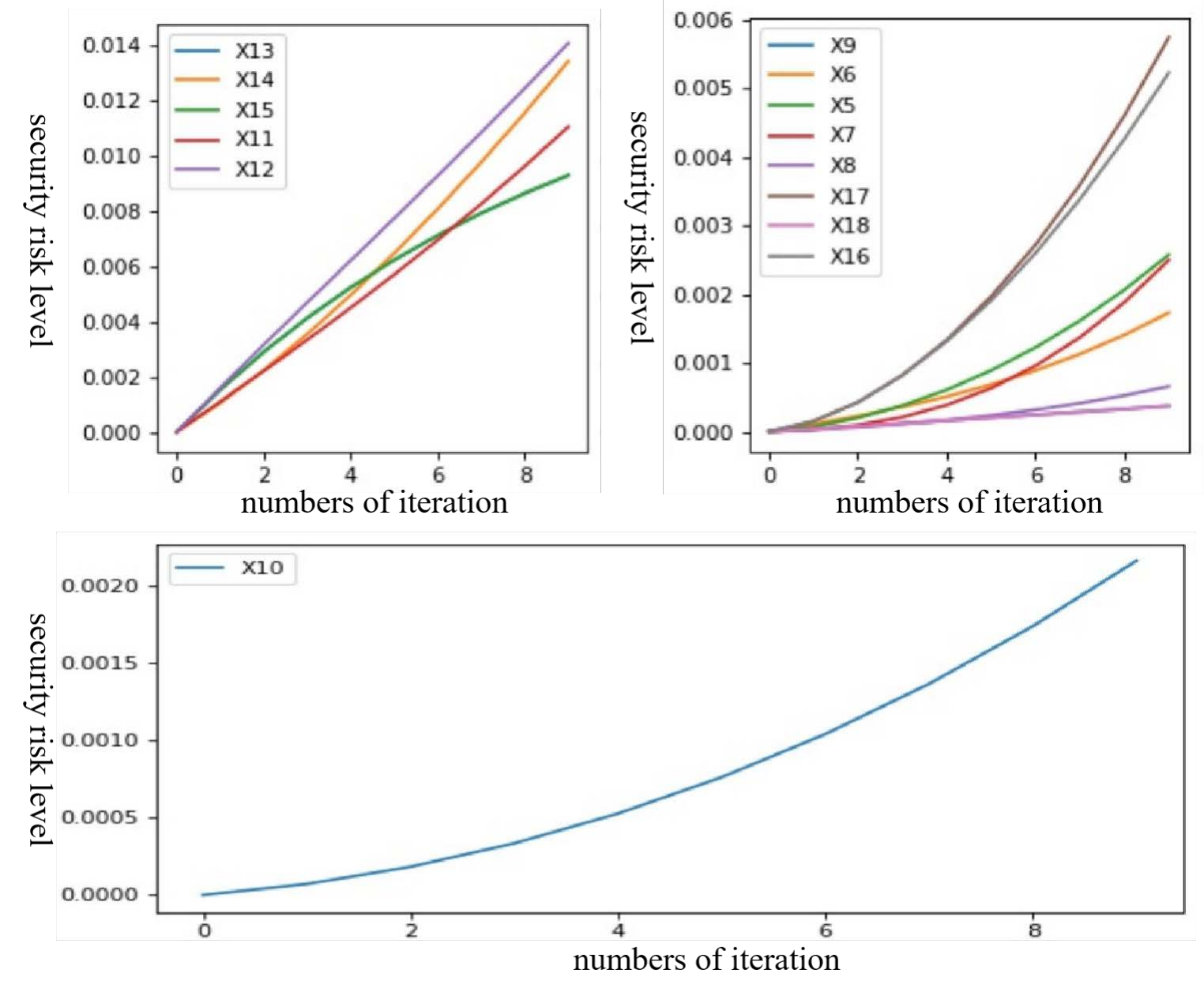}
		\caption{Risk propagation results}
		\label{Fig.5}
	    \end{figure}
     
Then, we explore the influence of security risk propagation on CTCS-3 components. After security risk propagating 10 iterations, we get the risk probability of each component, as shown in Fig. \ref{Fig.5}. We can see that the risk propagates faster in the central subsystem ($X11-15$) than in the trackside subsystem ($X5-9$, $X16-18$). And, with signal security communication, the security propagation is blocked slower than with TCP/IP communications, which is certified comparing to $X10$ and other components in the trackside subsystem. More specifically, for the trackside subsystem, the ranking of network risk sensitivity are: $X12(\text{Printer in CTC})>X14(\text{Workstation in CTC})>X11(\text{Server in CTC})>X15(\text{Interface device in CTC})>X13(\text{Remote maintenance router in CTC})$; For central subsystem, the ranking of network risk sensitivity are: $X17(\text{TSR})>X16(\text{Station CTC}) > X5(\text{RBC}) > X7 (\text{TCC})> X6 (\text{CBI})> X8 (\text{LEU})> X18(\text{GSM-R Center}) > X9(\text{On-board System})$.  The above assessment results reflect the real CTCS system characteristic that the central subsystem is more easier than trackside subsystem to propagate risks due to their communication attributes.
And this proved the correctness of the proposed DCAG-based method.

 \begin{table}[htbp]
\centering
\caption{Impact of network and wireless network attacks with different security risk levels}
\begin{tabular}{ccc}
\hline
risk level & wireless attack & network attack   \\ \hline
1          & 0.460142162     & 0.460142162    \\
2          & 0.461951800     & 0.460142166     \\
3          & 0.463755618     & 0.460142169    \\
4          & 0.465553633     & 0.460142172     \\
5          & 0.467345863     & 0.460142175     \\
6          & 0.469132323     & 0.460142178    \\
7          & 0.470913031     & 0.460142181    \\
8          & 0.472688004     & 0.460142184     \\
9          & 0.474457258     & 0.460142188     \\
10         & 0.476220810     & 0.460142191    \\ \hline
\end{tabular}
\label{Tab.8}
\end{table}

\begin{figure}[htbp]
		\centering
		\setlength{\belowcaptionskip}{-1em}
		\includegraphics[scale=0.32]{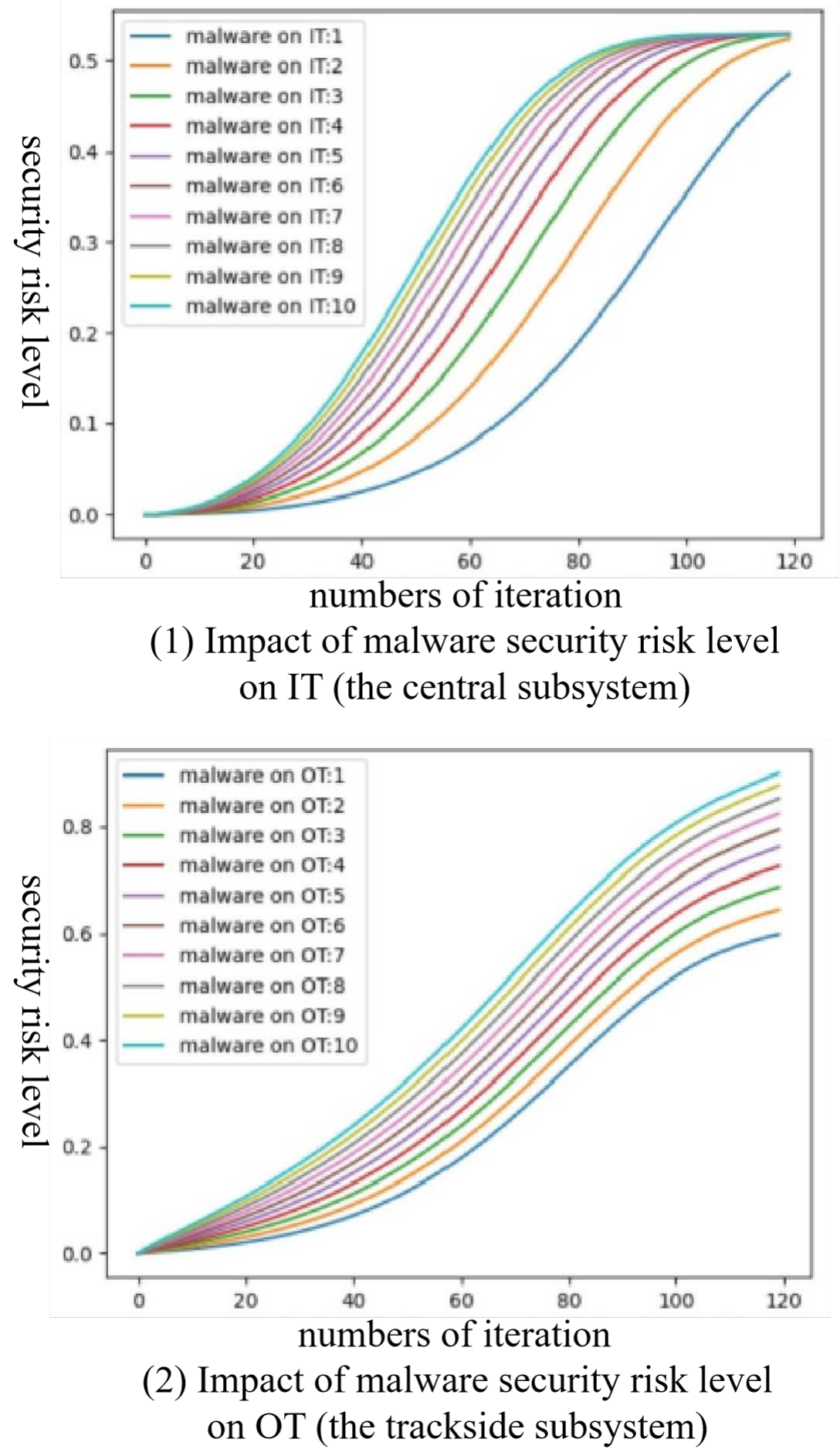}
		\caption{Impact of malware attacks with different security risk levels}
		\label{Fig.9}
	    \end{figure}
     
Furthermore, in order to determine which attacks within the system require additional resources for defense, we explore the influence of attacks with different security risk levels.  For networked attacks, we assign the relevant security risk levels from 1 to 10, individually. After 120 iterations of risk propagation, as the results are shown in Table \ref{Tab.8}, comparing wireless attacks and network attacks, we can see that wireless attacks have more influence than network attacks. 

For malware attacks, we assign the relevant security risk levels from 1 to 10, individually. After 120 iterations of risk propagation, the results are shown in Fig. \ref{Fig.9}. We can see that both malware attacks on the trackside subsystem and malware attacks on the central subsystem have more impact than the network attack and the wireless attack for the whole system. Comparing malware attacks on the trackside subsystem and the central subsystem, we can see that malware attacks on the trackside subsystem have more influence for the whole system. 

From the above simulations, we can infer some suggestions for the system defense strategies. 1) System assets on the central subsystem (IT) are more vulnerable to cyber attacks than systems assets on the trackside subsystem (OT). 2) Compared with network attacks on the central subsystem, more attention should be taken to wireless attacks on the trackside subsystem. 3) Compared with wireless/network attacks, malware attacks on both the central subsystem and trackside subsystems should be defended with more resources. 4) Malware attacks on the trackside subsystem have the greatest attack impact, which should be taken seriously.

\section{conclusion}
In this paper, we propose a security risk assessment framework based on the Dynamic Causal Attack Graph (DCAG) model for the train control system.  Firstly, we build the attack graph, and transfer it into the DCAG model with consideration of attack propagation and multi-stage attacks.
During the generation process, the root node of the graph is replaced from vulnerabilities to attack scenarios, the previous time slice nodes are added to deal with temporal security risk accumulation, and event causality relations are added on the edge to represent attack propagation dependencies. Then, we further propose the risk assessment method combining the Bayesian inference and logic gateway-based inference to solve the problems of temporal attack propagation and attack event causality propagation. With the DCAG model-based security assessment framework, we can not only appropriate security risk quantification calculations on CTCS-3 system, but also explore the importance of different attacks on system security risks, which is helpful to adjust the cyber security defense policy for CTCS-3 system. In the future, the security risk of other train control systems will be considered with the proposed DCAG-based security risk assessment framework.

\bibliographystyle{IEEEtran}
\bibliography{main.bbl}

\end{document}